\def\be{\begin{equation}}
\def\ee{\end{equation}}
\def\ber{\begin{eqnarray}}
\def\eer{\end{eqnarray}}
\def\bers{\begin{eqnarray*}}
\def\eers{\end{eqnarray*}}

\def\JPC{J. Phys. C}
\def\JPCM{J. Phys.: Condens. Matter}
\def\PR{{ Phys. Rev.}\ }
\def\PRL{{ Phys. Rev. Lett}\ }
\def\JPC{{ J. Phys. C: Solid State Phys}\ }
\def\JPCM{{ J. Phys.: Condens. Matter}\  }

\documentclass[aps,prb,twocolumn,groupedaddress,showpacs,amsmath,amssymb]{revtex4-1}
\usepackage{dcolumn}   
\usepackage{amsmath}
\usepackage{graphicx}    
\usepackage{subfigure}
\usepackage{color}

\newcommand{\comment}[1]{}
\usepackage{ifthen}
\newboolean{includefigs}
\setboolean{includefigs}{true}      
\newboolean{includetext}
\setboolean{includetext}{true}      
\newcommand{\condcomment}[2]{\ifthenelse{#1}{#2}{}}

\begin{document}

\title{Structural Properties and Relative Stability of (Meta)Stable Ordered, Partially-ordered and Disordered Al-Li Alloy Phases}

\author{Aftab Alam$^{1}$ and D. D. Johnson$^{1,2}$ }
\email[email: ]{aftab@ameslab.gov, ddj@ameslab.gov}
\affiliation{$^{1}$Division of Materials Science and Engineering, Ames Laboratory, Ames, Iowa 50011;}
\affiliation{$^{2}$Department of Materials Science \& Engineering, Iowa State University, Ames, Iowa 50011.}
\begin{abstract}
We resolve issues that have plagued reliable prediction of relative phase stability for solid-solutions and compounds.
Due to its commercially important phase diagram, we showcase Al-Li system because historically density-functional theory (DFT) results show large scatter and limited success in predicting the structural properties and stability of solid-solutions relative to ordered compounds.
Using recent advances in an optimal basis-set representation of the topology of electronic charge density (and, hence, atomic size), we present DFT results that agree reasonably well with all known experimental data for the structural properties and formation energies of ordered, off-stoichiometric partially-ordered and disordered alloys, opening the way for reliable study in complex alloys. 
\end{abstract} 
\date{\today}
\maketitle

\vspace{1cm}
\section{Introduction} 
Predicting alloy composition-temperature ($x~vs.~T$) phase diagrams has important practical implications, and much progress has been made in the last decade via \emph{ab-initio} study of phase stability.
Such predictions are a sensitive test as the relative stability of metallic alloys depend on small enthalpy differences. 
Notably, the order-disorder temperature of a superstructure is related to the difference between the formation enthalpy of the compound Al$_m$Li$_n$ ($x = n/(n+m)$) and mixing enthalpy\cite{PRL67.1779.1991} of disordered Al$_{1-x}$Li$_x$ alloy at the same composition $x$.\cite{Alam10} 
Historically, DFT results were sensitive to basis sets and approximations,\cite{Johnson-CMS8.54.1997} especially comparing ordering and mixing enthalpies from different codes.\cite{Mixing-EPL57.526.2002}
In particular, representations of charge densities are critical, especially for solid-solutions, where former ``bad'' results were not due, as often proposed, to the approximation to the disordered state.\cite{Alam09} 
How the disordered phases is addressed remains an ongoing issue, such as the longstanding discrepancies between approximations to partially-ordered (off-stoichiometric and thermal antisites\cite{Johnson2000.NiV}) and disordered phases, e.g., the coherent-potential approximation\cite{Johnson-CMS8.54.1997,Johnson-CPA,scr-cpa,Johnson2000.NiV} (CPA), special-quasi random structures\cite{Johnson-CMS8.54.1997,SQS1,SQS2} (SQS), and cluster expansions\cite{SluiterPRB42.10460.1990,Sluiter96,Johnson-CMS8.54.1997,Zarkevich2004.NiV} (CE).

{\par} Al-Li alloys, with their unusual elastic and structural properties and commercial importance,  have been the focus of many theoretical and experimental studies.
With doping to initiate precipitate growth in grains or grain boundaries, they are suitable materials for aerospace and automobile applications, due to their low density, high elastic modulus, and high strength-to-weight ratio. 
Between the stable endpoints of A1-Al (fcc) and A2-Li (bcc), numerous structural phases exist or compete,\cite{McAlister1990} e.g., the stable ${\beta}$ AlLi (B32), rhombohedral Al$_{2}$Li$_{3}$ (C33), monoclinic Al$_{4}$Li$_{9}$ (B2/m) and metastable (off-stoichiometric) ${\delta'}$-Al$_{3}$Li (L$1_2$). 
The ${\beta}$ AlLi, for example, is a promising candidate as an anodic material in high-energy density batteries; ${\delta'}$, which appears in the miscibility gap between Al and ${\beta}$, is used to precipitation-harden commercial alloys.\cite{precip-hard-Al3Li,expt-precip,Cocco1977} 
Substituting Li for Al not only makes the alloy less dense but increases unexpectedly the elastic moduli\cite{Noble82} even though the Young's modulus of Li (14 GPa) is seven times smaller than that of Al (91 GPa).
Also, the valence density leads to a bulk modulus of Al (83~GPa) five times larger than that of Li (15~GPa), giving rise to a sensitivity to basis sets if density is not represented properly.\cite{Alam09}

{\par} Comparing reliably properties of all competing phases in alloys is critical, and Al-Li is a sensitive, sufficiently complex, and yet unresolved case.
While there are some successes in describing the relative stability of the Al-Li ordered phases,\cite{Guo,Masuda-Jindo89,FS_Al3Li_2010} a reliable description for the disordered and partially-ordered phases is lacking. 
For Al-Li disordered phases, the virtual crystal approximation (VCA),\cite{CPA1} CPA,\cite{Korzhavyi-1994} and CE methods\cite{SluiterPRB42.10460.1990,Sluiter96} have been used to describe the $x$-dependent equilibrium volume and formation enthalpies, but quantitative prediction of the lattice constants and mixing enthalpy remains a problem.

{\par} Using an optimal site-centered basis for density and potentials,\cite{Alam09} we present \emph{ab initio} DFT calculations that compare reliably the phase stability (formation enthalpies) of competing ordered, partially-ordered, and disordered phases, and quantitatively reproduce (without adjustable parameters) the unusual alloying effects, including the lattice constants, i.e., $a$ vs. $x$, ill-described in the past.
We also  estimate the impurity formation energies at $x\simeq 0$ and $1$ (i.e., the solution enthalpies) of the A1-Al$_{1-x}$Li$_x$, a sensitive quantity due to differing electronic nature of Al and Li. 
For general configurations, we describe more properly the topology of electronic charge density, and hence atomic size, charge and alloying effects, especially in random alloys,\cite{Alam09,Alam10, Alam11} crucial in Al-Li for a reliable investigation of (meta)stable phases.\cite{Alam09}
We also resolve longstanding disagreements between various estimates of disordered energetics, e.g., CPA\cite{Johnson-CPA,scr-cpa} and SQS.\cite{Johnson-CMS8.54.1997}
To showcase the predictive accuracy, we compare and contrast our results to experimental data and those from linear augmented plane wave (LAPW),\cite{Guo} augmented spherical wave (ASW),\cite{Masuda-Jindo89} linear muffin-tin orbital (LMTO) methods, atomic sphere approximation (ASA).

\section{Computational Details}
{\par} We use a Korringa-Kohn-Rostoker (KKR) Coherent Potential Approximation (CPA) code \cite{MECCA} based on a weighted Voronoi polyhedra (VP)\cite{isoparam} basis defined from saddle-point radii (SPR) in the charge density.\cite{Alam09}
With this optimal basis-set, a proper representation of the topology of charge density and hence atomic sizes yields energy differences insensitive to basis-set L-truncation.\cite{Alam09,Alam10} 

{\par} We include {\emph{s-, p-, d-} and {\emph{f-}symmetries in the KKR Green's functions spherical harmonic basis, which is truncated at L$_\text{max}$=3, where L$\equiv(l,m)$.
Energy integration of Green's functions use a complex-energy Gauss--Chebyshev semicircular contour with 18 points.
The Brillioun zone integrations use Monkhorst and Pack \cite{Monkhorst} special k-point method using a $20\times20\times20$ mesh.
We use the von-Barth--Hedin\cite{von-Barth-1972} local density approximation (LDA) as parameterized by Moruzzi, Janak and Williams.\cite{MJW}
For random alloys, the screened-CPA\cite{scr-cpa} is used to incorporate metallic screening from charge correlations in the local chemical environment. 
More details are given elsewhere.\cite{Alam09,Alam10,Alam11}

{\par} The potential zero, i.e., {\it muffin-tin} {\it zero} $v_0$, can dramatically affect stability prediction for spherical potentials. 
We use a variational definition ($\text{X}$ stands for VP or ASA), 
\begin{equation}
v_0^{\text{X}} = \frac{\sum_{s} \int_{\text{MT}}^{\text{X}} \ d{\bf r}\ \rho_{s}^{\text{FP}}({\bf r}) \ V_{s}^{\text{FP}}({\bf r})}{\sum_{s} \int_{\text{MT}}^{\text{X}} d{\bf r} \ \rho_{s}^{\text{FP}}({\bf r})},
\label{eq1}
\end{equation}
with a sum over all sites in a unit cell.
We spherically average functions over the solid angle ($d\Omega=d\theta d\phi$) within $|{\bf r}| \le R_{\text{CS}}$, the maximal region required in a site-centered method. 
We then accurately integrate over the interstitial region of arbitrary VP (full shape) via isoparametric integration,\cite{isoparam} which is fast with error controllable to machine precision.

{\par}The total energies can be evaluate using weighted VPs, denoted by KKR-CPA(VP), or weighted-VPs approximated by unequal ASA spheres, denoted by KKR-CPA(ASA). 
For comparison, we also provide equal ASA sphere results - often used in other ASA-based codes.
Definition \eqref{eq1} yields kinetic energies that approach those of LAPW.\cite{Guo} 
For both ordered and disordered (i.e., CPA and SQS) results, we find significantly improved predictions using optimal SPR basis within each VP, in combination with the $v_0^{VP}$, even if the remainder of the calculations is based on ASA, as exemplified for Al-Li.
The CPA is applicable to an arbitrary $x$ (both solid-solutions and partially-ordered states\cite{Alam10}), and configurational averages can be enlarged via the DCA.\cite{2005Biava}


\begin{figure}[t]
\centering
\includegraphics[width=8.2cm]{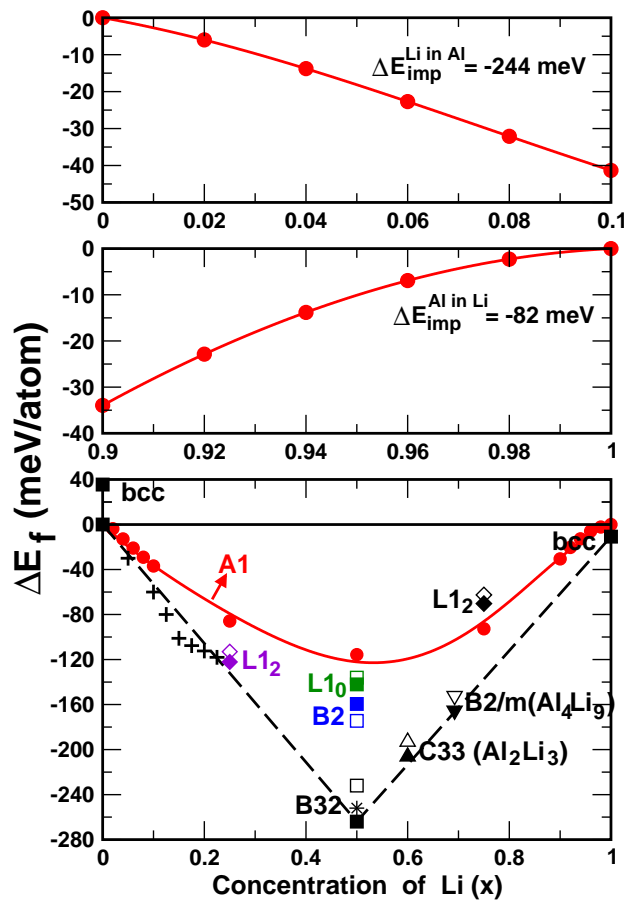}
\caption
{(Color online) (Bottom) $\Delta E_f$ ($m$eV/atom) vs. $x$ of A1 Al$_{1-x}$Li$_x$, and various fcc- and bcc-based compounds. 
KKR (LAPW\cite{Guo}) results are filled (hollow) symbols; B32 experiment\cite{Kubaschewski-1977} ($*$); A1 phase (solid curve); groundstate hull (dashed lines); and Al-rich (off-stoichiometric) L$1_2$ (pluses). 
Impurity formation energy $\Delta E_{\text{imp}}$ ($m$eV/atom) from a limit of a fit to $\Delta E_f$ (middle) Al in Li ($0.9<x<1.0$) and (top) Li in Al ($0<x<0.1$), values indicated in figure. }
\label{imp_form}
\end{figure}

\begin{table}[h!]
\caption{$\Delta E_f$ (in $m$eV/atom) for (dis)ordered alloys with equal and weighted-SPR-VP spheres relative to elements in same phase, using the variational $v_0$, Eq.~\eqref{eq1}. CPA and SQS results are from this work. CC denotes combined corrections. Other theoretical and experimental results are also given. 
\label{formation}}
\begin{ruledtabular}
\begin{tabular}{llcccl}
 System & ~~~~Method &   \multicolumn{2}{c}{Equal sphere}  &  \multicolumn{2}{c}{SPR sphere}  
\\
        &        & ${\Delta E}_{f}^{ord}$  & ${\Delta E}_{f}^{dis}$ & ${\Delta E}_{f}^{ord}$  &  ${\Delta E}_{f}^{dis}$  
\\
\hline
\scriptsize{A1-A2}
                  &  \scriptsize{KKR-ASA(VP)}                         &    &    &   $-35.5$  &   \\
\scriptsize{~~Al} &  \scriptsize{LAPW\cite{SluiterPRB42.10460.1990}} &    &    &  $-62.6$  & \\
          &  \scriptsize{EMTO\cite{Vitos1}} &    &    &  $-39.4$  & \\
          &  \scriptsize{KKR-ASA}                         &    &    &  $-19.0$   &   \\
\hline
\scriptsize{L1$_2$/A1}
  & \scriptsize{KKR-CPA(VP)}    &  $-78.3$  & $-4.60$   &  $-122.1$ & $-85.8$  \\
\scriptsize{~Al$_3$Li}
  & \scriptsize{KKR-SQS-16(VP)} &           & $-33.5$   &           & $-64.5$  \\
  & \scriptsize{LAPW\cite{SluiterPRB42.10460.1990}}&  $$       & $$       &  $-113.0  $ & $$  \\ 
  & \scriptsize{EMTO\cite{Vitos1}}&  $$       & $$       &  $-111.0  $ & $$  \\ 
  &  &  &  &  &  \\
  & \scriptsize{ASW-ASA\cite{Masuda-Jindo89}} &  $$       & $$       &  $~ -109.0^a  $ & $$  \\
  & \scriptsize{KKR-CPA(ASA)}   &  $-47.7$  & $-53.9$  &  $-132.8$ & $-91.1$  \\
  & \scriptsize{KKR-SQS-16(ASA)}    &           & $-55.1$   &           & $-55.0$  \\
  & \scriptsize{LMTO-ASA-CC\cite{CPA2}} &  $-75.6$       &~ n/a       &  $~ -133.7^b  $ &~~ n/a  \\
  & \scriptsize{LMTO-CPA-ASA\cite{CPA2}} &  $-57.4$       & $+91.1$       & $~ -282.9^c  $ & $-264.7^c$  \\
\hline
\scriptsize{L1$_0$/A1} 
  & \scriptsize{KKR-CPA(VP)}    &  $-71.8$  & $-54.2$  &  $-142.1$ & $-115.7$  \\
\scriptsize{~AlLi}
  & \scriptsize{KKR-SQS-8(VP)}  &           & $-55.7$   &           & $-116.0$  \\
  & \scriptsize{LAPW\cite{SluiterPRB42.10460.1990}} &  $$       & $$       &  $-139.5  $ & $$  \\
  & \scriptsize{EMTO\cite{Vitos1}}&  $$       & $$       &  $-156.8  $ & $$  \\ 
    &  &  &  &  &  \\
  & \scriptsize{KKR-CPA(ASA)}   &  $-58.3$  & $-32.3$  &  $-159.6$ & $-127.4$  \\
  & \scriptsize{KKR-SQS-8(ASA)}     &           & $-47.2$   &           & \, $-98.4$  \\
  & \scriptsize{LMTO-CPA-LDA\cite{Korzhavyi-1994}} &  $$       & $$       &  $$ & $-102.0$  \\
\hline
\scriptsize{B32/A2} 
  & \scriptsize{KKR-CPA(VP)}    &  $-187.8$  & $-36.3$  &  $-264.0$ & $-143.0$  \\
\scriptsize{~AlLi}
  & \scriptsize{KKR-SQS-8(VP)}  &           & $-65.7$   &           & $-121.0$  \\
  & \scriptsize{Expt\cite{Kubaschewski-1977}} &  $$       & $$       &  $-251.9 $ & $$  \\
  & \scriptsize{LAPW\cite{Guo}} &  $$       & $$       &  $-232.3  $ & $$  \\
  & \scriptsize{EMTO\cite{Vitos1}}&  $$       & $$       &  $-295.0  $ & $$  \\ 
    &  &  &  &  &  \\
  & \scriptsize{KKR-CPA(ASA)}   &  $-179.7$  & $-39.4$  &  $-279.6$ & $-142.9$  \\
  & \scriptsize{KKR-SQS-8(ASA)}     &           & $-67.2$   &           & $-152.4$  \\
\hline
\scriptsize{B2 AlLi}
  & \scriptsize{KKR-CPA(VP)}    &  $-118.2$  & $$       &  $-159.5$   & $$  \\
  & \scriptsize{LAPW\cite{SluiterPRB42.10460.1990}} &  $$       & $$       &  $-142.1  $ & $$  \\
  & \scriptsize{EMTO\cite{Vitos1}}&  $$       & $$       &  $-189.6  $ & $$  \\ 
    &  &  &  &  &  \\
    & \scriptsize{KKR-CPA(ASA)}   &  $-109.2$  & $$       &  $-138.5$   & $$  \\
  & \scriptsize{LMTO-ASA-CC\cite{Sluiter96}} &  $$       & $$       &  $-141.3  $ & $$  \\
  \hline
\scriptsize{L1$_2$/A1}
  & \scriptsize{KKR-CPA(VP)}    &     &     &  $-70.3$ & $-92.7$  \\
\scriptsize{~AlLi$_3$}    
  & \scriptsize{KKR-SQS-16(VP)} &     &     &         & $-78.3$  \\ 
  & \scriptsize{LAPW\cite{SluiterPRB42.10460.1990}} &  $$       & $$       &  $-68.0  $ & $$  \\
  & \scriptsize{EMTO\cite{Vitos1}}&  $$       & $$       &  $-91.3  $ & $$  \\ 
\hline 
\scriptsize{A1-A2}
                  &  \scriptsize{KKR-ASA(VP)}  &    &    &  $+8.2$  &   \\
\scriptsize{~~Li} & \scriptsize{LAPW\cite{Guo,SluiterPRB42.10460.1990}}     &    &    &    $-6.8$ &   \\
          &  \scriptsize{EMTO\cite{Vitos1}} &    &    &  $-3.9$  & \\
          &  \scriptsize{KKR-ASA}  &    &    &  $+4.1$  &   \\
\end{tabular}
\end{ruledtabular}
$^a$ Adjusted spheres via pseudopotential theory, see text. \\
$^b$ Charge-neutral, adjusted spheres with CC and L$_{max}=2$.\\
$^c$ Charge-neutral, adjusted spheres with L$_{max}=2$. 
\end{table}

\section{Results}  
We report formation and mixing enthalpies ($\Delta E_{f}$), impurity formation (or solution) energies ($\Delta E^{\text{x}}_{\text{imp}}$), and lattice constant ($a$) changes versus composition ($x$).
We address the critical charge representation in electronic-structure methods to make predictions from various theoretical techniques more consistent.
For Al-Li, comparing to experimental data should be done with care, as the literature contains large scatter in various values.
It is also well known that there is a Li deficiency that extends $0.3~mm$ into the bulk from the surface in Al-Li alloys.\cite{Ueda1985} 
Hence, experiments affected by this deficient region will not provide reliable data versus $x$.

\subsection{(Dis)Ordered Formation Energies}
Our results will clearly demonstrate that KKR-CPA(VP) correctly predicts the stability of ordered, disordered and partially-ordered phase all within the same fast code.\cite{MECCA} 
In Figure~\ref{imp_form} (bottom panel) we show the $\Delta E_f^{}$  versus $x$ for ordered Al$_m$Li$_n$ ($x=n/(n+m)$) and disordered Al$_{1-x}$Li$_x$ alloys.
Selected values of the results are provided in Table~\ref{formation}, and compared with LAPW and available experimental data, as well as ASA results, such as from linear and exact\cite{Vitos1} MTO (EMTO) methods.

{\par}Importantly, using an SPR optimal basis and Eq.~\eqref{eq1}, the KKR-CPA(VP) results for compounds are now in excellent agreement with LAPW results\cite{Guo} and measured data.\cite{Kubaschewski-1977} 
For example, the measured $\Delta E_f^{B32}$ is $-252\pm{10}~m$eV and our optimal-basis KKR(VP) result is $-264~m$eV, while LAPW finds $-232~m$eV. 
For metastable L1$_0$ AlLi, $\Delta E_f^{L1_0}$ is $-142~m$eV for KKR(VP) and $-140~m$eV for LAPW.
$\Delta E_f^{B2}$ for metastable B2 (CsCl) phase, predicted to exist at high pressure,\cite{Sluiter96} is  $-160~m$eV for KKR(VP) and $-142~m$eV for LAPW.

{\par} The KKR-CPA  mixing enthalpy $\Delta H_f^{A1}(x)$ for disordered fcc Al-Li is shown by the solid (red) curve in Fig.~\ref{imp_form}.
Generally, in phase diagram assessments, the solid-solution enthalpy of a fixed phase is fitted to a polynomial in $x$, such as,
\begin{equation}\label{eq-mixE}
     \Delta H_f^{A1}(x)=\sum_{m,n} L_{mn}\ x^{m} (1-x)^{n}\ \ \  \ \forall\ (m+n)\le 4 \ \  
\end{equation}
where $L_{mn}$ are the polynomial coefficients.
For A1 Al-Li, the asymmetry in the  $\Delta H_f^{A1}(x)$ in Fig.~\ref{imp_form} arises from the Madelung contributions to the total energy, which is proportional to  $(\Delta Q_i)^{2}$, the effective charge on a site, which also has a corresponding asymmetrical form as a function of $x$. 
Therefore, site charge is an important quantity in Al-Li, and a proper description of charge density and the corresponding transfer effect is necessary, as will be discussed below.

\subsubsection{(Off)Stoichiometric Al-rich L1$_2$}
{\par}The observed miscibility\cite{expt-precip} of L1$_2$-Al$_3$Li with respect to A1-Al and B32-AlLi is reproduced.
Indeed, the segregation line between pure Al and B32 sits at $-132~m$eV at $25\%$~Li, whereas L1$_2$ is $-122~m$eV, i.e., barely unstable but a very low-energy excitation; and, concomitantly, the stability of L1$_2$-Al$_3$Li relative to A1-Al$_{0.75}$Li$_{0.25}$ yields the observed L1$_2$ precipitation within the $\alpha(\text{Al}) + \beta$ two-phase field.\cite{expt-precip}
L1$_2$ remains metastable as free energies of the competing ordered phases change little because their order is near perfect whereas that of the Al-rich disordered phase decreases rapidly due to the increasing solubility of Li and the increasing configurational entropy. 

{\par}This argument can be supported by looking at the energetics for Al-rich, off-stoichiometric L$1_2$, shown as plus symbol in the lower panel of Fig. \ref{imp_form}, where the excess Al goes only to a Li corner-site and the three Al face-sites remain fully occupied by Al. 
These data indicate that a partially-disordered arrangement on the Al-rich side is more stable than the A1 phase, and, by comparing to the groundstate hull, slightly more favorable than $\alpha(\text{Al}) + \beta$ alone.
 Based only on 0~K enthalpy differences, we directly predict a higher stability of off-stoichiometric L1$_2$ between 8-20\%Li with a peak at $\sim$$15\%$, as observed\cite{Cocco1977} for the $\delta'(\text{Al$_{3}$Li})$ precipitates, where it undergoes a transformation to $\alpha(\text{Al}) + \delta(\text{AlLi})$ state at $\sim$$618~$K.\cite{Cocco1977}
At the $15\%$ peak, we estimated the maximum order-disorder temperature from $\Delta E_f^{\text{A1}-\delta`(\text{L1$_2$)}}$ as $581~$K.
Also, at perfect stoichiometry $\Delta E_f^{\text{A1-L1$_2$}}$ is $42~m$eV (or 488~K) and it compares well with the 522~K obtained via a much more involved CE fits using LAPW results,\cite{SluiterPRB42.10460.1990}
both of which should be compared to the measured eutectic temperature (869~K) showing the liquid phase is more stable than the homogeneous solid-solution.

\subsubsection{CPA vs. SQS}
{\par} For completeness, we compare the CPA to the SQS, within the same code and approximations.
The SQS is an ordered cell meant to approximate the two-site atomic correlations of a random alloy to a small cutoff, but it does not include configurational averaging as in the CPA, except as offered by the limited number of inequivalent sites within the ordered local environments.
In Table~\ref{formation}, we compared our SQS and CPA results at $25\%$ and $75\%$ (a $16$-atom cell\cite{SQS2}) and at $50\%$ (an average of two $8$-atom configurations\cite{SQS1}, one state having large positive formation energy and the other negative, as is typical).  
They agree from $2-20\%$ depending on composition $x$, but only if the SPR basis is used for both, as can be verified in Table~\ref{formation}; both CPA and SQS are sensitive to basis-representation but agree reasonably well within the optimal SPR basis, especially if larger SQS cells are used to improve the effective configurational average (see FCC 50\% case, e.g., in Table~\ref{formation}, where VP results shows much improved agreement).

\subsection{Impurity Formation Energies}
{\par} We now address the solution enthalpy (or impurity formation energy), a more sensitive quantity to the approximations used, because they are related to the slope of the formation energy in the impurity limits.
In the top and middle panel of Fig.~\ref{imp_form}, we show $\Delta E_f$ vs. $x$ for A1 phase in the two extreme limits ($0 \le x \le 0.1$ and  $0.9 \le x \le 1.0$).
The solution enthalpy must be determined from Eq.~\eqref{eq-mixE} using two restricted fits in the limits, i.e., for {B in A} ($0 \le x \le 0.1$) and vice-versa ($0.9 \le x \le 1.0$),
\begin{eqnarray}\label{eq-impE}
    \Delta E_{\text{f}}^{x \rightarrow 0(1)} &=& \sum_{m,n} L^{x \rightarrow 0(1)}_{mn}\ x^{m} (1-x)^{n} \\
    \Delta E_{\text{imp}}^{\text{B in A}} 
       &=& ~~\left[  \frac{\partial}{\partial x} \Delta E_{\text{f}}^{x \rightarrow 0}\right]_{x=0}  ;  
       \nonumber  \\      
    \Delta E_{\text{imp}}^{\text{A in B}}  
       &=& -\left[ \frac{\partial}{\partial x} \Delta E_{\text{f}}^{x \rightarrow 1}\right]_{x=1} \ . 
       \nonumber 
\end{eqnarray}
where $L^{x \rightarrow 0(1)}_{mn}$ are the restricted fitting coefficients.
The slopes at each endpoint are very different from one another and cannot be reliably extracted by a polynomial fit for all $x$ from Eq.~\eqref{eq-mixE}, as typically done.

{\par} Using Eq.~\eqref{eq-impE} and our KKR-CPA(VP) results, our $\Delta E_{\text{imp}}^{\text{Li in Al}}$  is $-244~m$eV/atom and  $\Delta E_{\text{imp}}^{\text{Al in Li}}$ is $-82~m$eV/atom .
A thermodynamic fit to experimental data using CALPHAD yields\cite{Mcalister82} $-222~m$eV/atom for $\Delta E_{\text{imp}}^{\text{Li in Al}}$ and $-154~m$eV/atom for $\Delta E_{\text{imp}}^{\text{Al in Li}}$. 
Our present results now agree much better with this one set of estimated experimental data, where notably only the Al-rich side has A1 measured data.
In the Li-rich end, we have ignored relaxation effects around the impurity, which leads to a decrease of $19~m$eV for the L1$_2$-AlLi$_3$ compound,\cite{Mixing-EPL57.526.2002} suggesting a potentially small increase in the magnitude of our $\Delta E_{\text{imp}}^{\text{Al in Li}}$. 

{\par}Midownik\cite{1986Midownik} has shown, however, that phase diagram fitting is not always able to predict structural energy differences accurately; hence, there is significant variation in the literature.
For example, for {Li in Al} and {Al in Li}, other CALPHAD fits yield $-131~m$eV/atom and $+972~m$eV/atom,\cite{chen89} respectively, the later being unphysical; and, another CALPHAD fit yields $-954~m$eV/atom and $-130~m$eV/atom,\cite{Saboungi1977} respectively, giving a phase diagram not in very good agreement with the experiment.\cite{Mcalister82}

{\par} Other theory results are in quantitative disagreement to experiment and the present results; for example, previous SQS results for $\Delta E_{\text{imp}}^{\text{Li in Al}}$ find $-309~m$eV/atom, and cluster expansions yields $-353~m$eV/atom, while former equal-sphere CPA results give $-358~m$eV/atom.\cite{Mixing-EPL57.526.2002}
All previous values were consistent, but none were obtained from proper impurity limit. 
Many of these discrepancies between results of the various electronic-structure based methods (CPA, CE, SQS, etc.) have been explained using Al-Ag alloys as case study,\cite{Johnson-CMS8.54.1997} where charge issues were not significant; whereas for Al-Li the charge representation is crucial, as also noted for ``big-atom/small-atom'' alloys.\cite{Alam10,isoparam}

\begin{figure}[]
\centering
\includegraphics[width=8.0cm]{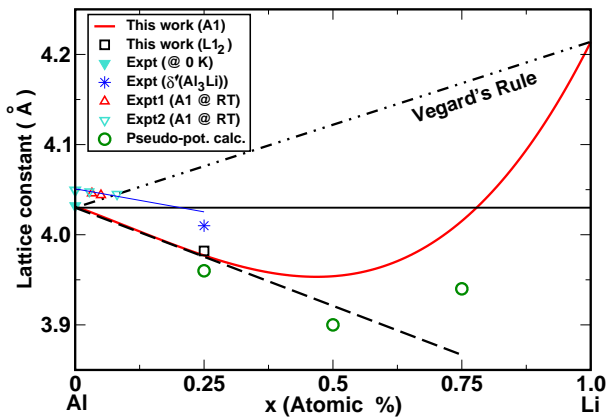}
\caption
{(Color online)  KKR-CPA(VP) $a$ vs. $x$ for A1-Al$_{1-x}$Li$_{x}$ [solid (red) curve], compared to other data, including  $a^{\text{Al}}_{\text{expt}}$  at 0~K.\cite{expt_pure-Al} $a^{A1}_{\text{expt}}$ from room temperature (RT) measurements [down\cite{expt1} and up\cite{exptAl,expt2} triangles]. $a_{\text{expt}}$  for $\delta'$(Al$_3$Li) precipitate at $523~$K [$*\cite{expt-precip}$]. Also shown are pseudo-potential results via first-order, perturbation theory [open circles].\cite{Masuda-Jindo89}}
\label{latt_cons}
\end{figure}

\subsection{Lattice Constants}
{\par} The minimized $a$ vs. $x$ for A1 Al$_{1-x}$Li$_x$ (solid curve) are shown in Fig.~\ref{latt_cons}, as well as for bulk L$1_2$ Al$_3$Li (square), with dilute limit slope added (dashed line) reproducing the well-known strong deviation to Vegard's rule (dot-dashed line).
Although the atomic volume of Li is $20\%$ larger than that of Al, addition of Li to Al initially causes a contraction of $a$ vs. $x$, which we also find in quantitative agreement with experiment. \cite{expt1,expt2,expt-precip}
Various experimental $a$'s reported\cite{expt1,expt2,exptAl,expt-precip} are
indicated by symbols, including Al-rich A1 and $\delta'$(Al$_3$Li) precipitates. 
First, our calculated $a$ for pure Al at 0 K is in agreement with experiment.\cite{expt_pure-Al}
Our results do not include thermal expansion effects, which are in room temperature measurements, explaining the $-0.7\%$ difference in calculated results.
Experimentally,\cite{expt-precip} $a$ for $\delta'$(Al$_3$Li) at R.T. shrinks by $1\%$ compared to pure Al, see Fig.~\ref{latt_cons}, and we find a reduction of $1.25\%$ at 0 K, in agreement with experiment (accounting for thermal expansion makes the agreement even closer).
Finally, a very sensitive measure is the slope of the $a$ vs. $x$  curve near aluminium, which is measured\cite{expt1,expt2,expt3} to be $-2.9$ to $-6.9\times10^{-5}~nm$ per at.\%Li.
The FLAPW impurity doping results\cite{SluiterPRB42.10460.1990} give $-5.1\times10^{-5}~nm$ per at.\%Li.
We find it to be $-11\times10^{-5}~nm$ per at.\%Li, within our CPA calculation.

\subsection{ASA versus VP}
For completeness, we provide the energetics from a wholly ASA implementation, including with $v_0^{\text{ASA}}$ in Eq.~\eqref{eq1} evaluated using ASA spherical (not VP) integrals, denote as KKR-CPA(ASA). 
Moreover, to better understand the sensitivity of the effective charge and Madelung energy, the calculations are done in two different ways:  (1) Standard equal atomic size (sphere or VP) for both Al and Li atoms (as done in many calculations), and (2) adjusting the sizes of each site by a weighted VP from the SPR found in a given environment.\cite{Alam09} 
From Table~~\ref{formation}, the SPR basis yields significantly improved energetics (especially compared to equal sphere ASA case) for both ordered and disordered phases, and eliminates the basis-set dependence for energy difference.\cite{Alam09} 
For instance, the SPR KKR-ASA result for ordered L$1_2$-Al$_3$Li $\Delta E_f^{L1_2}$ is $-133~m$eV/atom and it is in dramatic contrast to the equal-sphere KKR-ASA result of $-47.7~m$eV/atom. 
Improving the SPR-ASA basis using $v_0^{VP}$ and VP integrations (i.e., KKR-CPA(VP)) yields $\Delta E_f^{L1_2}=-122~m$eV/atom, in better agreement with $-113~m$eV/atom from LAPW.\cite{Guo}

\begin{table}[t]
\caption{Excess charges on the equal and weighted SPR of Li (first line) and Al (second line) atoms in (dis)ordered Al-Li.
\label{excess_charge}}
\begin{ruledtabular}
\begin{tabular}{llcccl}
 System & ~~~~Method &   \multicolumn{2}{c}{Equal}  &  \multicolumn{2}{c}{SPR}  
\\
        &        & ${\Delta Q}^{ord}$  & ${\Delta Q}^{dis}$ & ${\Delta Q}^{ord}$  &  ${\Delta Q}^{dis}$  
\\
\hline
\scriptsize{L1$_2$ Al$_3$Li}
  & \scriptsize{KKR-CPA(VP)}    &  $-0.450$ & $-0.291$ &  $-0.126$ & $-0.108$  \\
  &                            &  $+0.150$ & $+0.097$ &  $+0.042$ & $+0.036$  \\
  & \scriptsize{KKR-CPA(ASA)}   &  $-0.498$ & $-0.303$ &  $-0.141$ & $-0.132$  \\
  &                            &  $+0.166$ & $+0.101$ &  $+0.047$ & $+0.044$  \\
\hline
\scriptsize{L1$_0$  AlLi} 
  & \scriptsize{KKR-CPA(VP)}    &  $-0.101$ & $-0.069$ &  $-0.032$ & $-0.026$  \\
  &                            &  $+0.101$ & $+0.069$ &  $+0.032$ & $+0.026$  \\
  & \scriptsize{KKR-CPA(ASA)}   &  $-0.130$ & $-0.099$ &  $-0.082$ & $-0.058$  \\
  &                            &  $+0.130$ & $+0.099$ &  $+0.082$ & $+0.058$  \\
\hline
\scriptsize{B32 AlLi} 
  & \scriptsize{KKR-CPA(VP)}    &  $-0.203$ & $-0.187$ &  $-0.050$ & $-0.045$  \\
  &                            &  $+0.203$ & $+0.187$ &  $+0.050$ & $+0.045$  \\
  & \scriptsize{KKR-CPA(ASA)}   &  $-0.236$ & $-0.197$ &  $-0.098$ & $-0.105$  \\
  &                            &  $+0.236$ & $+0.197$ &  $+0.098$ & $+0.105$  \\
\hline
\scriptsize{B2 AlLi} 
  & \scriptsize{KKR-CPA(VP)}    &  $-0.080$ &          &  $-0.025$ &   \\
  &                            &  $+0.080$ &          &  $+0.025$ &   \\
  & \scriptsize{KKR-CPA(ASA)}   &  $-0.100$ &          &  $-0.068$ &   \\
  &                            &  $+0.100$ &          &  $+0.068$ &   \\
\end{tabular}
\end{ruledtabular}
\end{table}

\subsection{Charge Representations}
{\par} The phase stability and the bonding characteristics of the Al-Li alloys can be understood by looking at distribution of valence charge around each atom. 
A characteristic of this compound is that Li-atom redistributes some of its valence electron in between the Al bonds and the resultant strengthened Al bonds stabilize the compounds. 
The charge density topology in the Li-rich compounds is more crucial in stabilizing a particular phase.

{\par} In Table \ref{excess_charge}, the calculated local excess charges are provided within ASA and weighted-VP, for some ordered and disordered phases.  
From the data, there is clearly a  significant ``charge transfer'' from Li to Al for both ordered and disordered phases, as expected from electronegativity where Al (1.61) is larger than Li (0.98).
The excess (deficit) SPR-VP site charges from the KKR-CPA(VP), as expected, reflect a better charge neutrality than the SPR-ASA site charges more approximate KKR-CPA(ASA).
The KKR-CPA(VP) thus provides an improved estimate of the chemical potential and energetics, as shown, as well as charge effect due to the more proper charge density representation. 
In a VP-based calculations,  Madelung contribution to the total energy, as compared to an ASA, is minimized.

{\par} For the optimal SPR basis, the inscribed sphere reflect more appropriately the extent of the charge density on an atomic site, and hence a more reliable estimate of site charges; also the case for the unequal SPR-ASA results, but the energetics are sensitive to VP integrations.
For example, in AlLi for the SPR-VP basis, the excess (deficit) charge on Al (Li) remains almost unchanged for A2 and B32 phases (with only $0.005~e^-$ exchanged to Li from Al); whereas for the SPR-ASA sphere the charge remains small but is double that of the VP result.

{\par} For the equal-sphere ASA results,  however, there is an unphysical deficit charge on the smaller atom (Li) from the improperly described tails of the charge density of larger atom (Al), which have been arbitrarily cut off at the smaller radii (hence the dependence upon charge representation of the basis set we discussed).
The equal sphere results are unphysical, and the large charge exchange significantly impacts the total energy results, see discussion in Section~\ref{sec_discuss}, which is also responsible for the enormous variation in past DFT results.

\section{Discussion}  \label{sec_discuss}
Unlike other arbitrary (or unphysical) choices of charge density representation and basis set , the present SPR-based basis (weighted VP or ASA spheres) provides a unique, physics-based optimal representation of the charge topology for each atom type in a given environment.
This new basis reflects more properly the electronegativity (``charge transfer''), reduces the overlap error for ASA, and is valid for both ordered and disordered alloys, i.e., CPA and SQS approximations. 
For the sensitive case of Al-Li this has been demonstrated, which permits direct calculation of fully ordered, partially ordered and disordered alloys, as recently shown for the quantum criticality in the doped, intermetallic NbFe$_2$\cite{Alam11} and magnetic-storage alloys.\cite{Alam10}

{\par} Previously, Masuda-Jindo and Terakura\cite{Masuda-Jindo89} applied pseudo-potential-based, first-order perturbation theory\cite{Hafner-1987} to reveal a contractions in $a$ versus $x$ for Al-rich solid solutions, similar to experiment, which deviates from simple Vegard's rule due to effects from bulk moduli and atomic volumes.
They derived a relation for a mean ASA radii versus $x$ for the alloy, which was then used to perform a KKR-ASW study\cite{Masuda-Jindo89} of solid-solution hardening and softening using only ordered structures (Al$_7$Li and Al$_3$Li).
For L1$_2$-Al$_3$Li they found a decrease of $a$ by $1.5\%$ compared to pure Al.
In light of our present results, their adjustments to the ASA radii better reflects the electron density and ``charge transfer'' in agreement with our SPR-based VP and ASA.

{\par} Recently, Korzhavyi \emph{et al.}\cite{Korzhavyi-1994} performed LMTO-CPA-ASA calculations for the random alloy using charge corrections (similar to the scr-CPA) with an adjustable parameter, and \emph{a posteriori} correction for apparent charge transfer given by their choice of ASA spheres.
Only then did they find that the $a$ versus $x$ and mixing enthalpy started to agree qualitatively with experiment.
No such \emph{a posteriori} choices and corrections are needed for our optimal basis, uniquely chosen before any calculation.\cite{Alam09}

{\par} Finally, in an attempt to correct the apparent ``errors'' of the CPA, Singh and Gonis\cite{CPA2} proposed  ``charge-neutral'' spheres, an arbitrary, unphysical and, as recently discussed,\cite{Alam09}  unnecessary assignment. 
Their formation energy changed from positive to negative when the spheres radii were adjusted to give charge-neutrality, with the latter giving a clearly incorrect formation enthalpy of $-283$ 
($-265$)~$m$eV for the ordered (disordered) Al$_3$Li, as shown in Table \ref{formation}. 
For the ordered alloys, however, they were able to add \emph{combined-corrections} to the LMTO-ASA, which reduces ASA overlap errors due to the approximately 50\% increase in the Al ASA radii.
With combined-corrections the L1$_2$ formation enthalpy improves to $-134~m$eV; yet, they could not correct the CPA in a similar fashion.
Such an approach, however, does not represent the CPA charge density correctly.\cite{Alam09}
Simply put, for Al-Li, such a dramatic change in ASA radii to force charge neutrality arises solely because of the large difference in calculated bulk moduli (Al $0.72$~Mbar vs. Li $0.12$~Mbar) and volumes (Al $3.981~$\AA~and Li  $4.255~$\AA). 
Hence, only for ASA, the Al sphere will enlarge at the expense of Li because the Li sphere is easily compressed (the ratio of Al to Li bulk moduli is $6$), so the size of the Li sphere is expected to decrease much more than should be physically expected from the total density -- see original motivations discussed  in Ref.~\onlinecite{Masuda-Jindo89} and Ref.~\onlinecite{Alam09}.
So, when the moduli are substantially different, this adjustment does not describe the proper density, and adjusting spheres to minimize energy, say, is unwarranted.

{\par} Generally, however, using the saddle-point radii from the electronic density as the inscribed sphere of the VP (and using radical plane construction weighting to define full VP) offers a unique, physical and optimized representation, working for all ordered and disordered configurational averaging approximations, including the CPA.

\section{Summary}
For the commercially important Al-Li system, using a unique, physical and optimized representation of a site-centered charge density and potential in any local configuration, we accurately predicted, via KKR-CPA, the relative stability of all ordered, off-stoichiometric partially-ordered and disordered phases, including thermal antisites.\cite{Johnson2000.NiV}
We also resolved long-standing discrepancies throughout the literature for the sensitivity of Al-Li energy estimates for disorder alloys (CPA, SQS, and CE), permitting accurate prediction of formation enthalpies, solution enthalpies and structural properties for general alloy configurations, all within a single code as desired for a reliable study of phase stability in complex alloys.

\section{Acknowledgements} 
This work was supported by the U.S. Department of Energy, Office of Basic Energy Science, Division of Materials Science and Engineering (DE-FG02-03ER46026 - algorithm development) and Ames Laboratory (DE-AC02-07CH11358 - materials discovery applications). Ames Laboratory is operated for the U.S. DOE by Iowa State University.


\end{document}